\documentclass[fleqn,twoside]{article}
\usepackage{espcrc2}
\usepackage{graphicx}
\usepackage[figuresright]{rotating}

\newcommand{\AmS}{{\protect\the\textfont2
A\kern-.1667em\lower.5ex\hbox{M}\kern-.125emS}}

\newcommand{\beeq}{\begin{equation}}
\newcommand{\eneq}{\end{equation}}

\newcommand{\beeqa}{\begin{eqnarray}}
\newcommand{\eneqa}{\end{eqnarray}}

\title{
{\vspace{-1.2em} \parbox{\hsize}{\hbox to \hsize
{\hss  \normalsize UPRF-2004-20}}} \\
Exact supersymmetry on the lattice: the Wess-Zumino model}
\author{Marisa~Bonini\addressmark[parma]\address[parma]{Dipartimento di Fisica, Universit\`a di Parma and 
INFN Gruppo Collegato di Parma, Parco Area delle Scienze, 7/A, 43100 Parma, Italy} and 
Alessandra~Feo\addressmark[parma]\thanks{Talk given by A.~Feo}}
       
\begin{document}
\begin{abstract}
It is shown that the lattice Wess-Zumino model written in terms of Ginsparg-Wilson fermions is invariant under
a generalized supersymmetry transformation which is determined by an iterative procedure in the coupling constant. 
This transformation is non-linear in the scalar fields and depends on the superpotential parameters.
The implications of this lattice invariance are discussed.
\vspace{1pc}
\end{abstract}

\maketitle

\section{INTRODUCTION}
The major obstacle in formulating a supersymmetric theory on the lattice arises from the fact that
the supersymmetry algebra is actually an extension of the Poincar\'e algebra, which is explicitly
broken by the lattice.  
Indeed, in an interacting theory, translation invariance is broken since the Leibniz rule is not valid for
lattice derivatives \cite{dondi}. We know that Poincar\'e invariance is achieved automatically
in the continuum limit without fine tuning since operators that violate Poincar\'e invariance are all irrelevant.
A related issue is how to balance bosonic and fermionic degrees of freedom, as demanded
by supersymmetry: a naive regularization of fermions produces a wrong number of them. 
In order to construct a chiral lattice theory, without fermion doubling, the lattice Dirac operator 
$D$ should satisfy the Ginsparg-Wilson (GW) relation \cite{ginsparg}.

In this work we consider the four dimensional lattice Wess-Zumino model,
introduced in Refs.~\cite{fujikawa2,fujikawa}, that uses the GW fermion operator, 
and we show that it is actually possible to formulate the theory in a way that the full action is invariant
under a generalized supersymmetry transformation at a fixed lattice spacing. 
The action and the transformation are written in terms of the GW
operator and reduce to their continuum expression in the limit $a\to 0$.

\section{FOUR DIMENSIONAL WESS-ZUMINO MODEL ON THE LATTICE}
The action of the four dimensional Wess-Zumino model on the lattice has been introduced in 
Refs.~\cite{fujikawa2,fujikawa} and can be re-written as
\beeq
S_{WZ} = S_0 + S_{int} \, ,
\label{WZ}
\eneq
with
\beeqa
&&   S_{0} = \sum_x \bigg \{ \frac{1}{2} \bar \chi (1 - \frac{a}{2} D_1)^{-1} D_2 \chi \nonumber \\ 
&&  - \frac{1}{a} ( A D_1 A + B D_1 B)  
+ \frac{1}{2} F (1 - \frac{a}{2} D_1)^{-1} F \nonumber \\  
&&  + \frac{1}{2} G (1 - \frac{a}{2} D_1)^{-1} G \bigg\} \label{wz0} \, , \\[12 pt]
&&  S_{int} = \sum_x \bigg\{ \frac{1}{2} m \bar \chi \chi + m (F A + G B)  \nonumber \\  
&&  + \frac{1}{\sqrt{2}} g \bar \chi (A + i \gamma_5 B) \chi \nonumber \\
&&  + \frac{1}{\sqrt{2}} g \big[ F (A^2 - B^2) + 2 G (A B) \big] \bigg\} \, .
\label{wzint}
\eneqa
Here $A,B,F,G$ are scalar fields and $\chi$ is a Majorana fermion that satisfies the Majorana condition, 
$\bar \chi = \chi^T C$, where $C$ is the charge conjugation matrix, with  
$C^T = -C$ and $C C^\dagger = 1$. 
Our conventions are $ C \gamma_\mu C^{-1} = - (\gamma_\mu)^T $ and $ C \gamma_5 C^{-1} = (\gamma_5)^T $.
The operators $D_1$ and $D_2\equiv  \gamma_\mu D_{2 \mu}$ entering in the kinetic terms of the Wess-Zumino
action are related to the Neuberger operator $D$, given in \cite{neuberger}, by 
\beeq
D_1 = \frac{1}{4} \mbox{Tr}(D) \, , \qquad D_2 = \frac{1}{4} \mbox{Tr}(\gamma_\mu D) \, .
\eneq
Notice that, in terms of $D_1$ and $D_2$ the GW relation  becomes
\beeq
D_1^2 - D_2^2 = \frac{2}{a} D_1 \, .
\label{gw1}
\eneq
The action (\ref{WZ}) reduces to the continuum Wess-Zumino one in the limit $a \to 0$. 

\subsection{The lattice supersymmetry transformation}
As discussed in \cite{fujikawa}, $S_0$ is invariant under superymmetry, while 
$S_{int}$ breaks this symmetry  because of the 
failure of the Leibniz rule at finite lattice spacing \cite{dondi}. 
One expects that this invariance can be restored by adding irrelevant terms  to the action
(\ref{WZ}). 

In \cite{bonini} it has been shown that the full action (\ref{WZ}) is invariant under 
the following supersymmetry transformation 
\beeqa
&& \delta A = \bar \varepsilon \chi = \bar \chi \varepsilon  \nonumber \\
&& \delta B = -i \bar \varepsilon \gamma_5 \chi = -i \bar \chi \gamma_5 \varepsilon \nonumber \\
&& \delta \chi = - D_2 (A - i \gamma_5 B) \varepsilon - (F - i \gamma_5 G) \varepsilon + 
g R \varepsilon \nonumber \\ 
&& \delta F = \bar \varepsilon D_2 \chi \nonumber \\
&& \delta G = i \bar \varepsilon D_2 \gamma_5 \chi \, ,
\label{complete}
\eneqa
where $R$ is a function depending on the scalar and auxiliary fields and their derivatives
and can be determined in perturbation theory. 
By expanding $R$ in powers of $g$ 
\beeq
R = R^{(1)} + g R^{(2)} + \cdots
\label{expansion}
\eneq
and imposing the symmetry condition order by order in perturbation theory, we find
\beeq
R^{(1)} = ((1 - \frac{a}{2} D_1)^{-1} D_2 + m )^{-1} \Delta L 
\label{r1}
\eneq
with 
\beeqa
&& \hspace{-0.6cm} \Delta L \equiv  \frac{1}{\sqrt{2}} \Big\{2 (A D_2 A - B D_2 B) - D_2 (A^2 - B^2) \nonumber \\
&&  + 2 i \gamma_5 \Big[(A D_2 B + B D_2 A) - D_2 (A B)\Big] \Big\} 
\eneqa
and for $n \geq 2$
\beeqa
&& \hspace{-0.6cm} R^{(n)} = - \sqrt{2} ((1 - \frac{a}{2} D_1)^{-1} D_2  + m)^{-1} (A + i \gamma_5 B) \nonumber \\
&& \phantom{R^{(n)} = }  \times R^{(n-1)}  \, .
\label{rn}
\eneqa
Thus, the function $R$ is non-linear in the fields and can be seen to be the formal solution of 
\beeqa
&& \hspace{-0.8cm} \big[(1 - \frac{a}{2} D_1)^{-1} D_2  + m  + \sqrt{2} g (A + i \gamma_5 B) \big] R = 
\Delta L \, . \nonumber 
\eneqa
Notice that, for $a \to 0$ the transformation (\ref{complete}) reduces to the continuum
supersymmetry transformation, since $\Delta L$ vanishes in this limit. Indeed, $\Delta L$ is different from zero because
of the breaking of the Leibniz rule for a finite lattice spacing.

\subsection{The algebra}
We now study the algebra associated to the lattice supersymmetry transformation (\ref{complete}).
In particular, carrying out the commutator of two supersymmetries we must find a transformation which is still a 
symmetry of the Wess-Zumino action, i.e. the transformations of the fields form a closed algebra, order
by order in $g$. Here we explicitly check this fact up to order $g^1$, even though 
the calculation can be generalized to any order.
The commutator of two supersymmetries on the scalar field $A$ is \cite{bonini}
\beeqa
&& \hspace{-0.7 cm} [\delta_2,\delta_1] A = -2 \bar \varepsilon_1 \gamma_\mu \varepsilon_2 \Big\{  D_{2 \mu} A \nonumber \\ 
&&  + \frac{g}{\sqrt{2}}  \frac{m(1 - \frac{a}{2} D_1)}{m^2(1 - \frac{a}{2} D_1) + \frac{2}{a} D_1} 
\Big[D_{2 \mu}  (A^2 - B^2) \nonumber \\ 
&& - 2 (A D_{2 \mu} A - B D_{2 \mu} B) \Big] \Big\} \,.
\label{delta12A}
\eneqa
Similar expressions can be found for the other scalar fields.
For the fermion field $\chi$ a Fierz rearrangement is needed and after some algebra we find \cite{bonini}
\beeqa
&& \hspace{-0.7 cm} [\delta_2,\delta_1] \chi = -2 \bar \varepsilon_1 \gamma_\mu \varepsilon_2 \Big\{ D_{2 \mu} \chi \nonumber \\
&&  \hspace{-0.7 cm} - \frac{g}{\sqrt{2}} \frac{m(1 - \frac{a}{2} D_1) - D_2}{m^2(1 - \frac{a}{2} D_1) + 
\frac{2}{a} D_1} \bigg( D_2 (A - i \gamma_5 B) \gamma_\mu \chi \nonumber \\ 
&& \hspace{-0.7 cm} + (A + i \gamma_5 B) D_2 \gamma_\mu \chi 
 - D_2 \big[ (A - i \gamma_5 B) \gamma_\mu \chi \big] \bigg) \bigg\}  \, . \nonumber
\label{delta12chi}
\eneqa
Therefore, the general expression of the commutator of two transformations is 
\beeq
[\delta_1 , \delta_2] \Phi = \alpha^\mu P^\Phi_\mu(\Phi) \, , \quad \Phi = (A,B,F,G,\chi) \, ,
\eneq
where $\alpha^\mu = -2 \bar \varepsilon_2 \gamma^\mu \varepsilon_2$ and $P^\Phi_\mu(\Phi)$ are polynomials in 
$\Phi$ defined as 
\beeq
P^\Phi_\mu(\Phi) = D_{2 \mu}\Phi + O(g) 
\label{deltaalpha}
\eneq
where the order $g^1$ contributions  can be read in (\ref{delta12A}) for the case $\Phi=A$. 
We have verified that the closure works, i.e. the action (\ref{WZ}) is invariant under the transformation 
\beeq
\Phi \to \Phi + \alpha^\mu P^\Phi_\mu(\Phi) 
\label{deltaalpha1}
\eneq
up to terms of order $g^1$. 
Notice that, in the continuum limit, $D_{2 \mu} \to \partial_\mu $  and the transformation (\ref{deltaalpha1}) 
reduces to 
$\Phi \to \Phi + \alpha^\mu \partial_\mu \Phi$,
since terms of order $g^1$  in (\ref{delta12A}) vanish due to the restoration 
of the Leibniz rule as $a \to 0$.

\subsection{Ward-Takahashi identity}
The existence of this exact symmetry should be responsible for the restoration of supersymmetry 
in the continuum limit. The study of the Ward-Takahashi identity (WTi) associated to the exact lattice
supersymmetry we have introduced can be done by generalizing the analysis performed by 
Golterman and Petcher in \cite{golterman} for the two dimensional Wess-Zumino model. 
By making a variation of the generating functional with respect to the lattice supersymmetry transformation 
(\ref{complete}), the WTi reads $< J_\Phi \delta \Phi > = 0$, 
where $J_\Phi $ are the sources for the fields $\Phi = (A,B,F,G,\chi)$. 

As an example we explicitly verify the simple WTi
\beeq
< F > - i \gamma_5 < G > - g < R > = 0 
\label{ward}
\eneq
to the order $g^1$. 
From the action (\ref{WZ}), the free propagators are 
\beeqa
&& \hspace{-0.4cm} < A A >^{(0)} = < B B >^{(0)} = -{\cal M}^{-1} (1 - \frac{a}{2} D_1)^{-1} \nonumber \\
&& \hspace{-0.4cm} < F F >^{(0)} = < G G >^{(0)} = \frac{2}{a}{\cal M}^{-1}  D_1  \nonumber \\
&& \hspace{-0.4cm} < A F >^{(0)} = < B G >^{(0)} = m \,{\cal M}^{-1} \nonumber \\
&& \hspace{-0.4cm} < \chi \bar \chi >^{(0)} = -{\cal M}^{-1} ((1 - \frac{a}{2} D_1)^{-1} D_2 - m) \nonumber \, ,
\eneqa
where 
\beeq
{\cal M} = \frac{2}{a} D_1 (1 - \frac{a}{2} D_1)^{-1} + m^2 \, . 
\eneq
Then using the standard perturbative expansion we find 
\beeqa
<F(k)>^{(1)} &=&  \frac{\delta^4(k)}{m} \int_{p} {\cal M}^{-1}(p) \nonumber \\
&&  \hspace{-0.4cm} \times \mbox{Tr} \bigg[ (1 - \frac{a}{2} D_1(p))^{-1} D_2(p) - m \bigg]\nonumber \\
&&  \hspace{-0.4cm} + 4  \delta^4(k) \int_{p} {\cal M}^{-1}(p) = 0 \, .
\label{f}
\eneqa
while the one point function of $G$ is zero at this order.
Finally, the last term of the WTi (\ref{ward}),
\beeqa
&&  \hspace{-0.6cm} < R >^{(0)} = ((1 - \frac{a}{2} D_1)^{-1} D_2 + m)^{-1} <\Delta L>^{(0)} \, , \nonumber 
\eneqa
vanishes since $D_2(p)$ is an odd function of the integration momentum $p$.
Notice that, even if each term of the WTi (\ref{ward}) is zero separately, 
there are non-trivial cancellations between the fermionic and scalar terms in (\ref{f}).

\section{CONCLUSIONS}
We have shown that it is actually possible to
formulate the theory in such a way that the full action is invariant under a lattice supersymmetry
transformation at a fixed lattice spacing.
This supersymmetry transformation introduces a function $R$ which is non-linear in the scalar fields
and depends on the parameters $m$ and $g$ entering in the $S_{int}$.
We believe that the lattice supersymmetry we have introduced automatically leads to a restoration of 
continuum supersymmetry without additional fine tuning. We are currently investigating on this issue.
Obviously the most important question is whether these ideas may be extended to supersymmetric 
gauge theories where one expects that the GW relation plays an important role in the construction
of a lattice supersymmetry.

\end{document}